\documentclass[a4paper,nocompress]{spie}  %>>> use for US letter paper
\usepackage[]{graphicx}

\title{A broadband scalar optical vortex coronagraph} 

\author{Ronny Errmann\supit{a,b}, Stefano Minardi\supit{a}, and Thomas Pertsch\supit{a} 
\skiplinehalf
\supit{a}Institute of Applied Physics, Abbe Center of Photonics, Friedrich-Schiller-Universit\"at Jena, Max-Wien-Platz 1, 07743 Jena, Germany; \\
\supit{b}Astrophysical Institute and University Observatory, Friedrich-Schiller-Universit\"at Jena, Schillerg\"a{\ss}chen 2-3, 07745 Jena, Germany\\
}

\authorinfo{Further author information: S.M.: E-mail: stefano@stefanominardi.eu, Telephone: +49 (0)3641 947 848\\  R.E.: E-mail: Ronny.Errmann@uni-jena.de, Telephone: +49 (0)3641 947 518}
\begin{document} 
  \maketitle 

%%%%%%%%%%%%%%%%%%%%%%%%%%%%%%%%%%%%%%%%%%%%%%%%%%%%%%%%%%%%% 
\begin{abstract}
In recent years, new coronagraphic schemes have been proposed, the most promising being the optical vortex phase mask coronagraphs. In our work, a new scheme of broadband optical scalar vortex coronagraph is proposed and characterized experimentally in the laboratory. Our setup employs a pair of computer generated phase gratings (one of them containing a singularity) to control the chromatic dispersion of phase plates and achieves a constant peak-to-peak attenuation below 1:1000 over a bandwidth of 120\,nm centered at 700\,nm. An inner working angle of $\lambda$/D is demonstrated along with a raw contrast of 11.5\,magnitudes at 2$\lambda$/D. A more compact setup achieves a peak-to-peak attenuation below 1:1000 over a bandwidth of 60\,nm with the other results remaining the same.
\end{abstract}

%>>>> Include a list of keywords after the abstract 


%%%%%%%%%%%%%%%%%%%%%%%%%%%%%%%%%%%%%%%%%%%%%%%%%%%%%%%%%%%%%
\section{INTRODUCTION}
\label{sec:intro}  % \label{} allows reference to this section
Broadband coronagraphy with deep nulling and small inner working angle has the potential of delivering images and spectra of exoplanets and other faint objects located near a much brighter source. Recently, attention was driven to the possibility to use phase rather than amplitude masks to suppress efficiently the star light while retaining an inner working angle close to the diffraction limit\cite{rod97}.
Following this intuition, many different types of phase mask coronagraphs have been suggested and experimented in the last decade\cite{maw12}, the common denominator being that the phase mask is used to scatter the light of the central star outside the pupil aperture and thus can be easily removed by a circular aperture used as Lyot stop. Particularly interesting are the so called vortex phase mask coronagraphs\cite{foo05}, which transform the starlight beam into an optical vortex, a ring shaped beam with a null in the center and a spiral wavefront. Most of the proposed phase mask coronagraphs suffer however from strong chromatism, due to the fact that a specific phase shift cannot be generated by a single mask for all wavelengths simultaneously\cite{swa05}. As a result, scalar phase mask coronagraphs can attain high rejection of the central star only if operated with small bandwidths, a fact which is not compatible with the requirements of spectroscopic analysis of faint companions.
A solutions to the problem of chromatism has been proposed\cite{maw05} and consists in generating a polarization optical vortex by exploiting the form birefringence of sub-wavelength gratings (so called vectorial vortex coronagraphs, VVC). A potential disadvantage of VVC is however that, contrary to scalar vortex coronagraphs, polarization states have to be split to perform focal-plane adaptive optics\cite{mal95} for active reduction of speckle noise\cite{maw10}.

Here we propose an alternative solution for the chromaticity of scalar vortex coronagraphy in the visible band by use of a pair of computer generated phase gratings (one of them containing a singularity) to control the chromatic dispersion of phase plates. The proposed scheme, being inherently scalar, solves also the problem of polarization splitting affecting VVC, when used in combination with focal-plane adaptive optics. %We used a setup, consisting in general of a unit to simulate the telescope pupil, the vortex generation unit and the actual coronagraph. For the vortex generation, which uses the phase gratings, two setups were tested. %, first with a spatial filter is inserted between the holograms to select only their first diffraction order\cite{err13}, and second a compact setup with attached phase gratings by their substrate sites. 
%We present the performance of the setups, like peak-to-peak attenuation as well as the inner working angle of the coronagraph, as well as comparing the properties of both.
%
%The paper is organized as follows: in section~2 optical vortices in general as well as the implementation in for coronagraphy are discussed, while we present our setups of the coronagraph in section~3. Our results are presented in section~4, followed by the summary, in which we discuss further improvements as for example using active optics.

The paper is organized as follows: in Section~2 optical vortices as well as their use for coronagraphy are discussed. We present our setup of the coronagraph in Section~3. The set-up consists in general of a unit to simulate the telescope pupil, the vortex generation unit and the actual coronagraph. For the vortex generation unit, which uses the phase gratings, two setups were tested. Section~4 presents the performance of our setups, like peak-to-peak attenuation as well as the inner working angle of the coronagraph, as well as comparing the properties of both setups. The summary, in which we discuss further improvements, is following in Section~5.

\section{OPTICAL VORTICES AND CORONAGRAPHY}
\begin{figure}[t]
   \begin{center}
   \begin{tabular}{c}
   \includegraphics[width=0.7\textwidth]{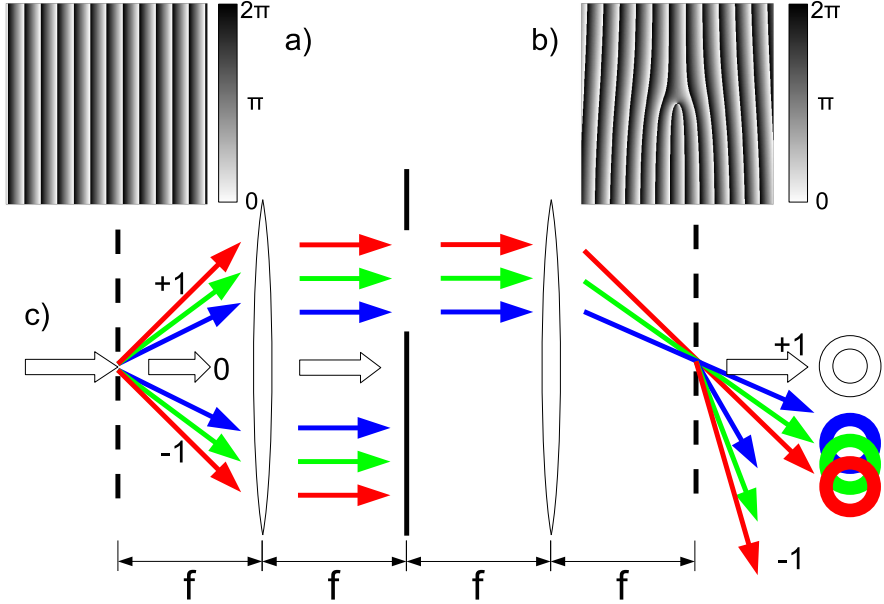}
   \end{tabular}
   \end{center}
   \caption[Vortex scheme] 
   { \label{fig:vortexscheme} 
Cartoon illustrating the generation of broadband optical vortex by means of off-axis, fork holograms. The unit consists of a grating (a), spatial filter ©, and the vortex phase plate (grating with singularity m=2) (b). In our implementation we used lenses with $f=100$\,mm.}	%Period ???
   \end{figure}

Optical vortices are fields exhibiting a helical wavefront with a central phase singularity and a coincident null amplitude point. 
The circulation of the phase gradient around the singularity is $m2\pi$, with the integer $m$ denoting the topological charge. 
With monochromatic light, optical vortices are usually generated by means of spiral phase masks\cite{foo05} or forked holograms\cite{Vasnetsov} (see Fig.~\ref{fig:vortexscheme}.b)  which imprint the vortex phase profile on the incoming light beam. 

For astronomical applications, optical vortices with even topological charge are particularly interesting.
In fact, if the center of the vortex phase mask is placed exactly at the focus of a diffraction limited, clear aperture telescope (that is the light beam has the shape of an Airy disk),  the modified beam will exhibit a ring-like structure in the pupil plane (the so called 'ring of fire'). The inner part of the ring of fire has exactly 0 amplitude up to the entrance pupil radius $R_\mathrm{P}$ and all the light is scattered in the outside region \cite{maw05, swa09}. This fact can be exploited in coronagraphs to attenuate the light of a bright star with a Lyot stop placed in the pupil plane and  shaped as a circular aperture of radius $R_\mathrm{P}$.
Indeed, a source (\textit{e.g.} a faint companion) focused with an offset from the center of the vortex phase mask won't be transformed into an optical vortex and will pass through the aperture of the Lyot stop with negligible losses.

The problem with spiral phase masks is that their operation is limited to quasi-monochromatic light, as the phase delay for a fixed optical path difference is wavelength dependent \cite{swa05}. Vortex generation with off-axis, fork holograms is also troubled by the fact that the angle of diffraction of the vortex beam is dependent on wavelength, as expected from any grating-like photonic structure.  

The latter problem can however be lifted if the fork hologram is illuminated by angularly dispersed light, as demonstrated by Bezuhanov et al. with a setup suitable for vortex generation with broadband laser pulses\cite{bez04}. 
In our work\cite{err13},  we analyzed the performance of this particular scheme of broadband vortex generation scheme for astronomical coronagraphy.

The broadband vortex generation setup is shown in Fig.~\ref{fig:vortexscheme}.c and consists of a grating (G, Fig.~\ref{fig:vortexscheme}.a), a spatial filter and a computer-generated, fork hologram (CGH, Fig.~\ref{fig:vortexscheme}.b).  
The function of the grating G is to pre-disperse the white light beam in order to compensate for the dispersion of the CGH. To this end, the period of the grating G is matched with that of the fork hologram CGH and the transverse magnification of the spatial filter is 1.
The spatial filter is necessary to select only the first diffraction order from the grating and avoid obtaining a superposition of optical vortices of different topological charge at the output of the setup (for a hologram, the topological charge of the vortex beam is given by $m\cdot n$, where $m$ is the intrinsic topological charge of the hologram and $n$ is the diffraction order under consideration). 
A superposition of different vortices would lead to a distortion of the ring of fire and a consequent reduction of the attenuation of the light of the central star by the coronagraph.

\section{LABORATORY SETUP}
The main layout of the laboratory setup of our vortex coronagraph demonstrator consists in 3 distinct units, namely 1) the Airy disk pattern generator (simulates a diffraction limited, clear aperture telescope), 2) the vortex generation unit, and 3) a spatial filter including the Lyot stop and the imaging system based on a 8 bit CCD. 

\begin{figure}[t]
   \begin{center}
   \begin{tabular}{c}
   \includegraphics[width=0.8\textwidth]{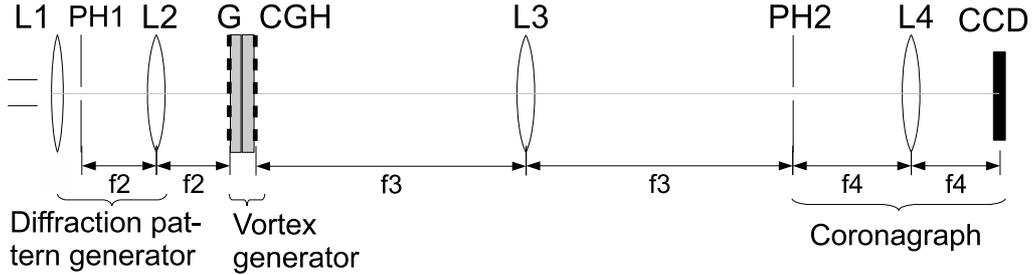}
   \end{tabular}
   \end{center}
   \caption[Setup] 
   { \label{fig:optical_setup} 
Optical setup of the new coronagraph (setup B) with compact vortex generator. In the original setup (setup A) the vortex generation unit consists of the components shown in Fig.~\ref{fig:vortexscheme}c. The lenses L1 ($f=8$\,mm) and L2 ($f=40$\,mm), and the pinhole PH1 ($D=75\,\mu$m) create the Airy disk of the fiber to simulate the beam of a telescope. The beam passes through the grating (G), directly followed by the CGH, which creates the optical vortex. Both phase plates were attached by their substrate sides. Lens L3 ($f=400$\,mm) images the beam on the pinhole PH2 ($D=700\,\mu$m), which acts as Lyot stop and blocks the ring of fire. A pupil transformation with lens L4 ($f=75$\,mm) transforms the image of PH2 on the CCD.}
   \end{figure}

The Airy disk, as the diffraction pattern of a circular aperture, is created by imaging light from a single-mode fiber to a pinhole (PH1, see Fig.~\ref{fig:optical_setup}). A lens (L2, Fig.~\ref{fig:optical_setup}) is then used to make a pupil transformation of the pin-hole aperture and project the Airy disk pattern at the input of the vortex generation unit.
As a light source, we injected in the fiber light either from a HeNe (633\,nm) or from a programmable white light source which could emit in the wavelength range from 640\,nm to 780\,nm. 

In our experiments we used two distinct setups for the vortex generation unit, hereafter indicated as setup A and setup B.
Both vortex generation schemes employed a phase grating and a fork phase hologram which were blazed to diffract more than 90\% of the light in the first order of diffraction at the design wavelength of $\lambda_0=650$\,nm. The fork phase hologram has intrinsic topological charge $m=2$.
Both elements were manufactured at the Fraunhofer Institute Jena by replicating on a thin polymer layer ($10\,\mu$m) a master hologram obtained by grey tone laser lithography. The polymer layer was spin coated on a 1mm-thin glass substrate. The period of both gratings was chosen to be $g=50 \mu$m to ease the fabrication process and avoid the 
effects of polarization-dependent diffraction efficiency.

In setup A, the vortex generation scheme was implemented according to the scheme outlined in Fig.~\ref{fig:vortexscheme}.c.
For the spatial filter we used two $f=100$ mm achromatic doublet lenses with broadband antireflection coating for the R- and I- bands (Thorlabs). An iris blind was used to select the first diffraction order of the grating G. The efficiency of the transmission through this unit is 75\% over a band width of 160\,nm\cite{err13}. This includes losses of $\sim16$\%, as the hologram and grating were not coated with antireflection layer causing a reflectivity of each surface of $\sim4$\%.
The fork hologram CGH was mounted on x-y micrometric translation stage to align precisely the beam to the vortex phase singularity. 

In setup B, the vortex generation unit consists only of the grating and CGH joined together by their substrate sides, 
as shown in Fig.~\ref{fig:optical_setup}.
The choice of this scheme is dictated mainly by the request to have a compact setup for the vortex generation which is compatible with existing coronagraphic instruments. The elimination of the spatial filter is justified by the fact that the diffraction efficiency of the gratings exceed 90\% at the design wavelength, so that the contribution of higher diffraction orders to the white light vortex beam are negligible, at least in a narrow band around to the design wavelength. With this solution, the setup was shortened by 40\,cm ($4f$ from Fig.~\ref{fig:vortexscheme}c), respect to setup A, making the mounting of the coronagraph on a telescope much easier, as mass and impact of flexure is reduced. 
The thickness of the combined phase plate (2 mm) is much smaller than the diffraction length of the illuminating Airy disk. 
To align the Airy disk precisely on the phase singularity, the combined CGH and grating were mounted on an x-y micrometric translation stage.

The third part of the coronagraphic demonstrator (the Lyot spatial filter) consists of a $\sim700\,\mu$m pinhole acting as Lyot stop, which is positioned at the pupil plane of lens L3. The light passing through the pinhole is projected on a CCD camera after passing through a lens L4 performing the final pupil transformation. To obtain high dynamic range images with our 8 bit CCD camera, several images of different exposure time are taken, which are then scaled by their exposure time and combined by their weighted initial flux.

\section{Results}

\begin{figure}
   \begin{center}
   \begin{tabular}{c @{\hspace{0mm}} c}
   \includegraphics[width=0.49\textwidth]{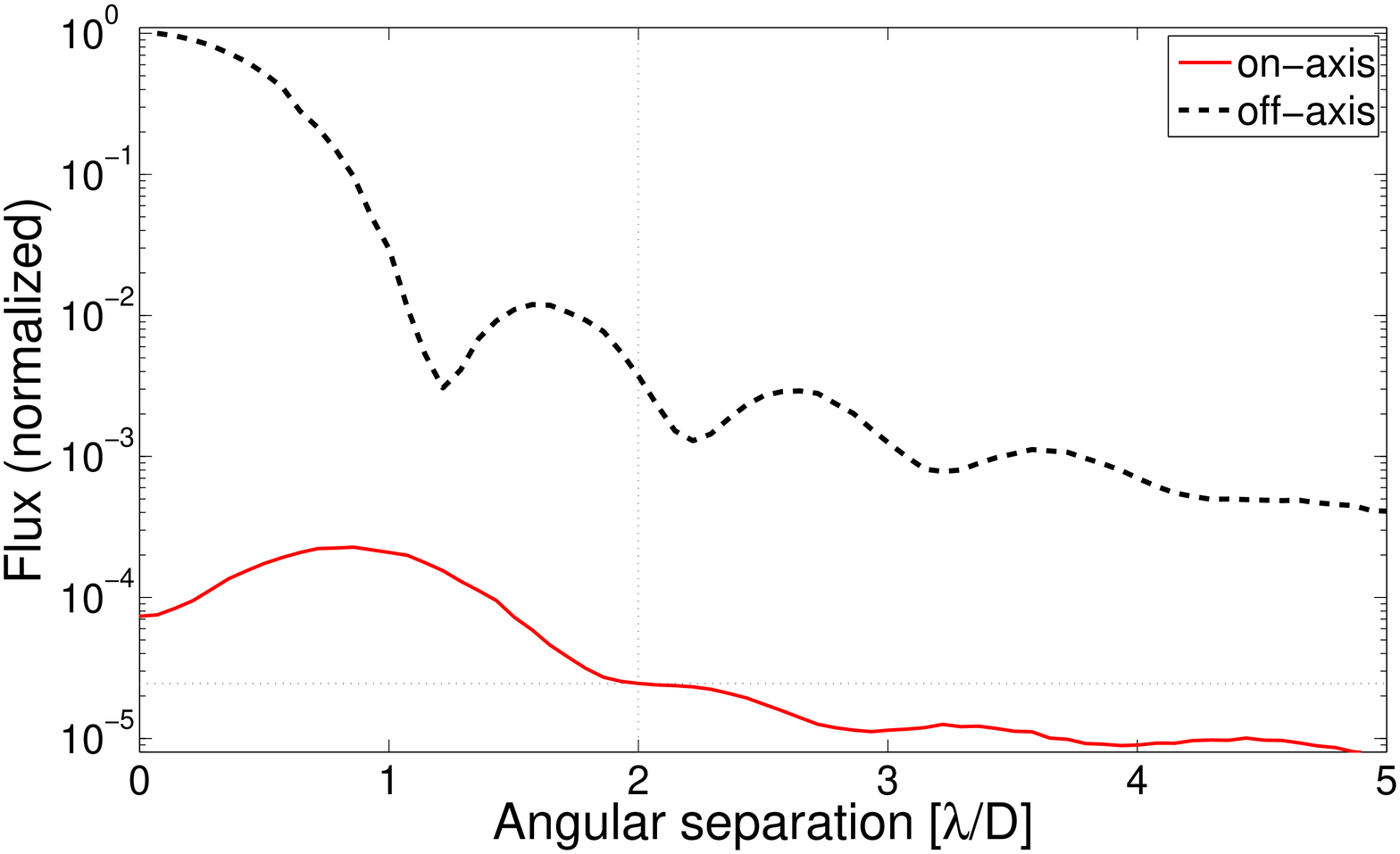} & \includegraphics[width=0.49\textwidth]{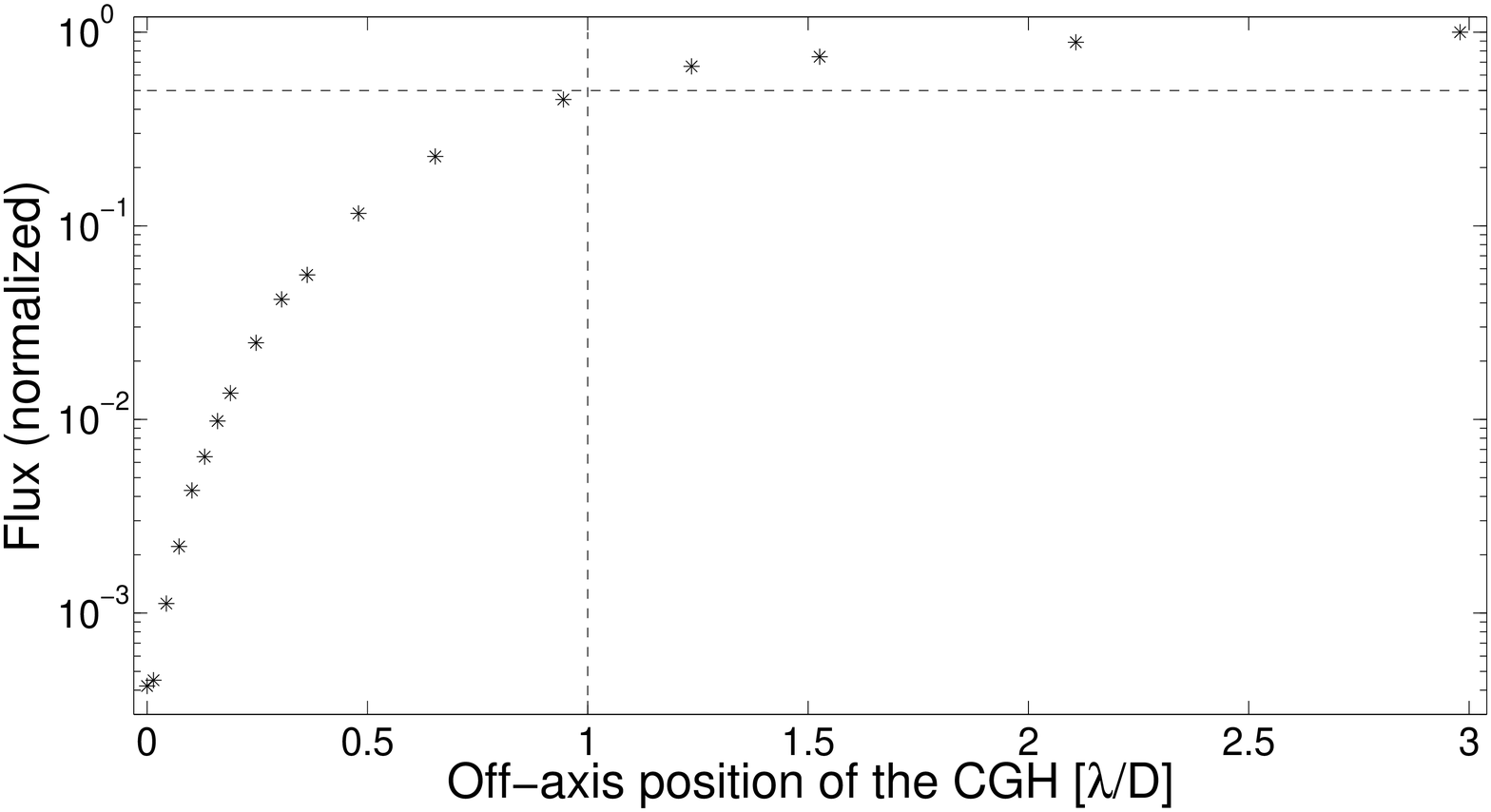}
   \end{tabular}
   \end{center}
   \caption[Radial profile of the Flux and Off axis contrast] 
   { \label{fig:radial_profile-offaxisflux} 
Left: azimuthally integrated point spread function profiles for fork phase plate on- and off-axis using the broad-band light source ($\Delta\lambda=120$\,nm, centered at $\lambda=700$\,nm). At $2\lambda/D$ the flux is attenuated to $2.5~10^{-5}$, meaning a contrast of 11.5\,mag could be reached.
Right: The flux of the brightest pixel as function of the angular separation between beam and phase singularity in the CGH, normalized to the flux at separation $3 \lambda/D$. Light from a HeNe laser was used. The dashed lines indicate attenuation of 50\% and separation of 1$\lambda/D$.}
   \end{figure} 

As an initial test of the vortex coronagraph (setup A), we measured the attenuation of the setup with a broadband light source ($\Delta\lambda=120$\,nm).
The result is illustrated in Fig.~\ref{fig:radial_profile-offaxisflux} (left). The unattenuated/attenuated point spread functions (PSF) were obtained by (1) centering the phase singularity on the optical axis of the instrument ('on-axis' curve), hence blocking the ring of fire with PH2, and (2) shifting the phase singularity off-axis by several $\lambda/D$ so that the artificial star is transmitted through PH2 ('off-axis' curve).
The peak-to-peak attenuation of the coronagraph is $3\cdot10^{-5}$ while at $2\lambda/D$ the flux is suppressed by $2.5\cdot10^{-5}$ (corresponding to 11.5 magnetudes). For greater separations the flux converges to a constant value about $10^{-5}$ weaker than the peak of the unattenuated PSF. This noise floor is probably due to a combination of detector noise and coherent speckle noise. The latter one is caused by aberrations of our optical components\cite{mal95} and could be suppressed with focal-plane adaptive optics.

The transmission of the setup A depending on the angular separation between the Airy disk and the phase singularity is shown in Fig.~\ref{fig:radial_profile-offaxisflux} (right). The diagram demonstrate how the CGH affects a close companion. As expected from $m=2$ vortex coronagraphs, at an angular separation of $\lambda/D$ about 50\% of the flux is supressed while at separations $\ge2\lambda/D$ a companion is nearly unaffected.

%\begin{figure}
%   \begin{center}
%   \begin{tabular}{c}
%   \includegraphics[width=0.8\textwidth]{off-axis-contrast.eps}
%   \end{tabular}
%   \end{center}
%   \caption[Off axis contrast] 
%   { \label{fig:offaxisflux} 
%The flux of the brightest pixel as function of the angular separation between beam and phase singularity in the CGH, normalized to the flux at separation $3 \lambda/D$. Light from a HeNe laser was used. The dashed lines indicate attenuation of 50\% and separation of 1$\lambda/D$.}
%   \end{figure} 

\begin{figure}
   \begin{center}
   \begin{tabular}{c}
   \includegraphics[width=0.8\textwidth]{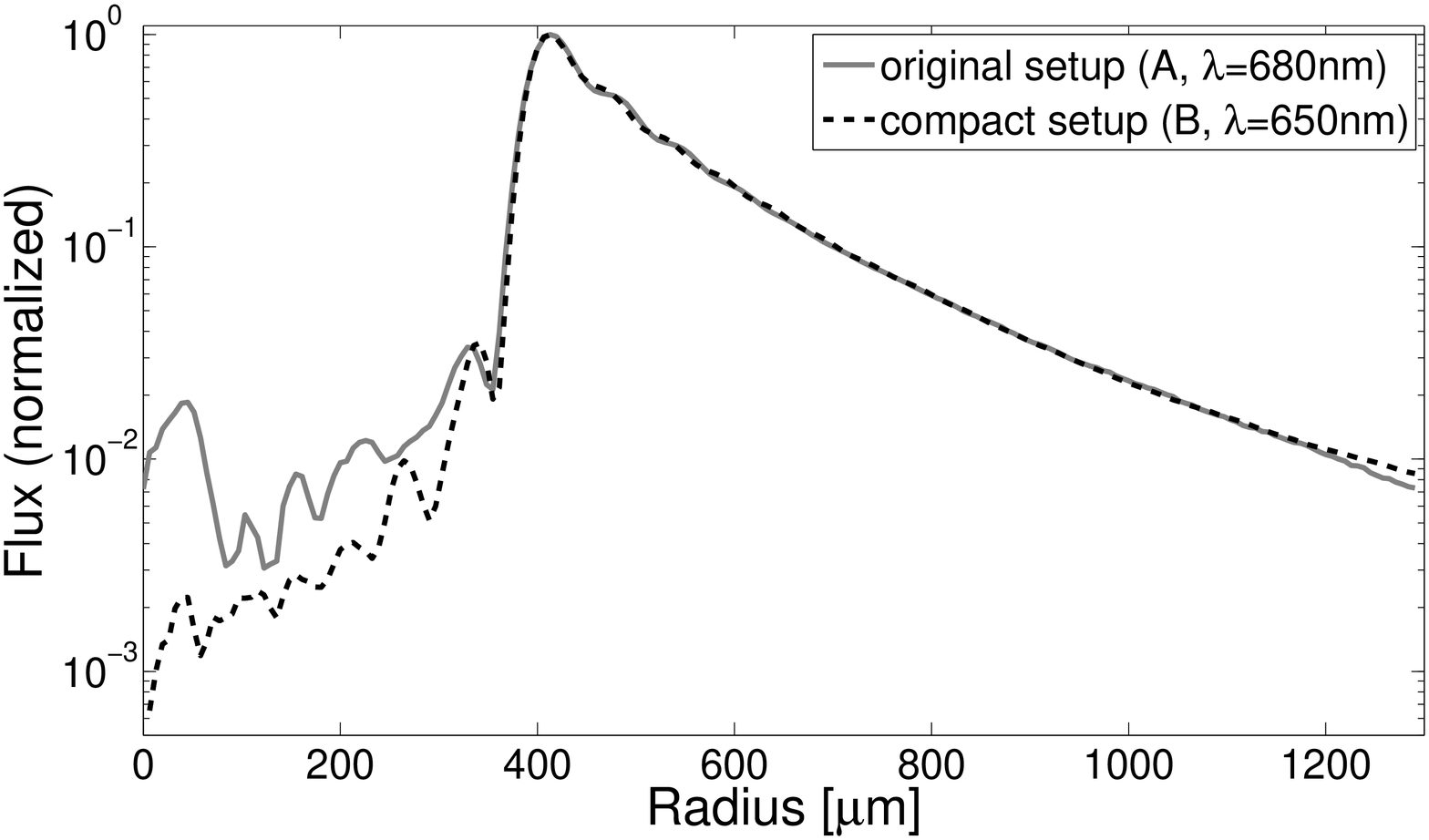}
   \end{tabular}
   \end{center}
   \caption[Rings of fire] 
   { \label{fig:rof_compare} 
Comparison between the radial profiles of the ring of fire between setup A (additional lenses and spatial filter between both phase plates, gray solid line) and setup B (black dashed line). The bandwidth of the light was $\Delta\lambda=5$\,nm for both measurements.}
   \end{figure}

We next compared the performance of setup A and B.
Fig.~\ref{fig:rof_compare} show the radial profiles of the rings of fire of both setups A and B. We imaged the ring of fire by locating the CCD camera in the pupil plane of lens L3, at the position of the Lyot stop PH2. In both cases, the area inside the ring of fire is not completely darkened due to low order aberrations of the setup. The slightly deeper values inside the ring of the compact setup is most probably an artifact due to the better dynamic range of the images taken with setup B.

\begin{figure}
   \begin{center}
   \begin{tabular}{c}
   \includegraphics[width=0.8\textwidth]{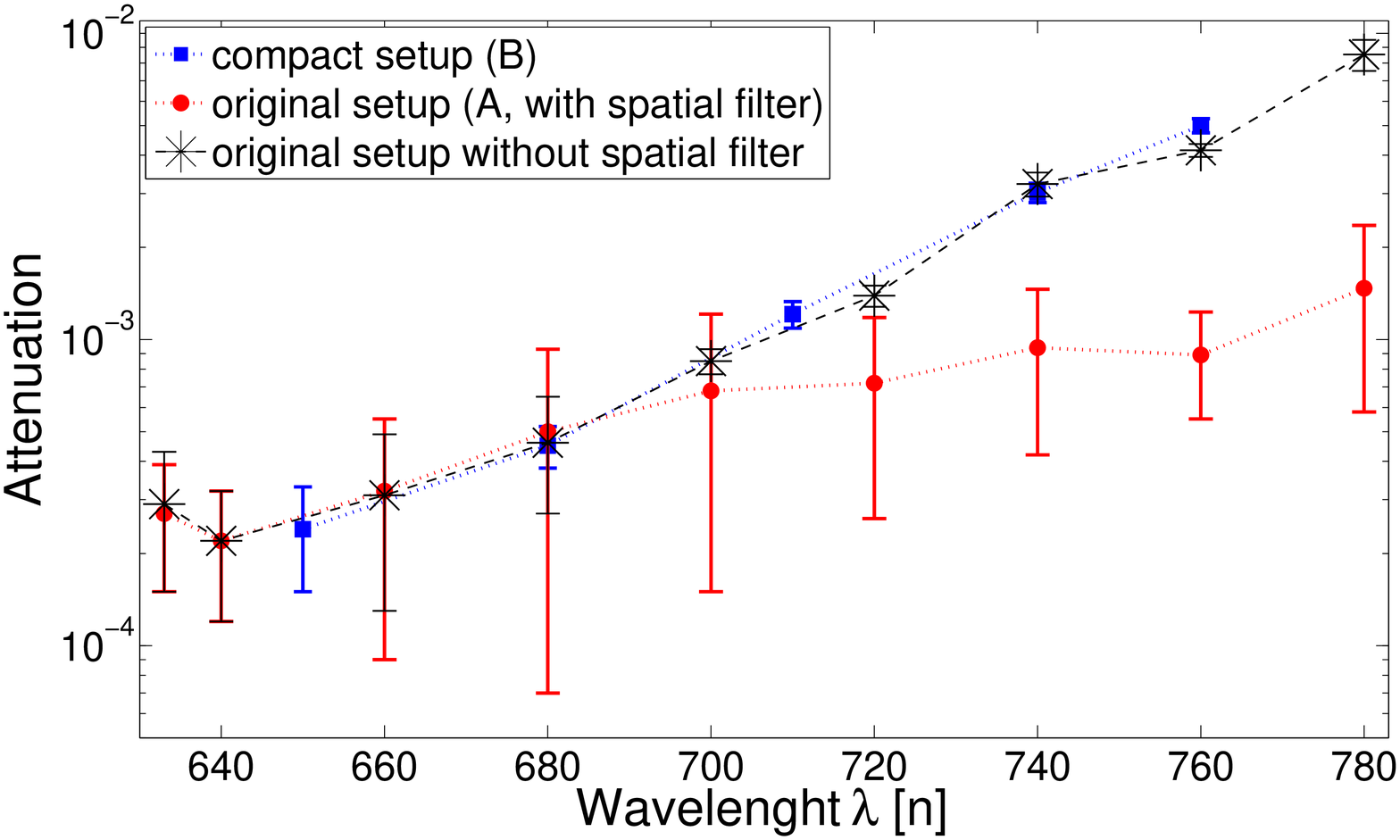}
   \end{tabular}
   \end{center}
   \caption[Attenuations] 
   { \label{fig:attenuation_compare} 
Peak-to-peak attenuations as measured for the two setups. Blue squares: compact setup (B), as shown in Fig.~\ref{fig:optical_setup}, red circles: initial setup A\cite{err13}, black stars: initial setup without the blend between grating and CGH. The best performance could be reached close to the design wavelength of the phase plates of $650$\,nm.}
   \end{figure}

Finally, we compared the attenuation of both coronagraphic setups. As mentioned before, the attenuation was measured as the fraction of the maximum values in the azimuthal integrated PSF in the attenuated and unattenuated case, respectively.
%Thereby, the attenuation was measured as the fraction of the brightness between two setups of the CGH: (1) centering the Airy disk of the white light source on the phase singularity, and (2) putting the Airy disk far off axis of the CGH. 
% In the second case, the light is unaffected by the phase singularity and passes only two times the grating. In the astronomical application the attenuation gives the fraction of light to what a star centered on the singularity is faded. 
Fig.~\ref{fig:attenuation_compare} shows the measured peak-to-peak attenuation as a function of the wavelength for setup A with spatial filter mask (red circles), setup A without spatial filter mask (black stars), and setup B (blue squares). Close to the design wavelength of 650\,nm the three curves are overlapped irrespective of the filtering of the higher diffraction orders of the grating ($n\not=+1$). For a wavelength detuning $\Delta\lambda\leq50$\,nm (8\% bandwidth relative to the carrier frequency) from the design wavelength the differences between the filtered and unfiltered settings are not significant.
For a wavelength detuning of 100\,nm (15\% relative bandwidth) and higher, the attenuation of the filtered setup A is 25\% of the other two settings. This behavior clearly is caused by the effect of the unfiltered higher diffraction orders, as demonstrated by the fact that the contrast of the unfiltered setup A  is the same as that of setup B.

%The decreasing attenuation with increasing $\Delta\lambda$ toward the construction wavelength is caused by???. defocusing, aberation?

\section{SUMMARY}
We have shown that we could reach a high contrast over a large bandwidth with the setup of a broadband optical vortex coronagraph, using scalar phase masks. These high contrasts are necessary to do spectroscopy on faint objects, \textit{e.g.} atmospheres of exoplanets.

Two different setups were tested, one (A) with a spatial filter used to remove stray light from higher diffraction orders of the grating, and a compact setup (B) without spatial filtering. Both setups reach peak-to-peak attenuations below 0.1\% over a significantly large bandwidth. 
Setup (B) exhibits  a contrast below 0.1\% over a bandwidth of 8\% (640 to 700\,nm), smaller as compared to the bandwidth of 17\% measured for setup A. This effect is related to the reduction of the diffraction efficiency in the first order for light detuned from the design wavelength of the phase plates. We were limited with our white light source to wavelength $\geq640$\,nm, hence in principle a larger wavelength range with reasonable contrast is expected. We estimate a wavelength range of 620 to 700\,nm (12\%) for the compact setup. This range is enough to cover transmission bands of the atmosphere.
The lower performance of the setup B can be traded-off with its compactness and compatibility to existing coronagraphic setups. 

The throughput of the vortex generation unit is about 75\%. To increase the efficiency of the instrument, CGH and grating could be written to the same substrate plate and could be coated with an antireflective layer on all components. We expect the throughput to be 90\% for this case.

Advantage of the tested setup is the insensitivity to polarization state of light and the scalability to longer wavelengths (near- or mid-infrared atmospheric transmission bands). 
As the brightness contrast between planets and stars decreases toward longer wavelengths, the coronagraph can then be used for direct imaging and spectral investigation of extrasolar planets. Assuming that we can achieve the same efficiency of the broadband scalar vortex coronagraph in the mid-infrared, we notice that a peak-to-peak attenuation below $\approx 0.1$\% in combination with a camera dynamics of 10$^4$ and image processing techniques\cite{laf07} would allow to detect objects as faint as 10$^{-6}-10^{-7}$ ($\Delta m=15-17$ magnitudes) compared to the central star, thus already permitting to image directly old exoplanets. 

We notice that the achievement of higher contrasts with our setup is not excluded, considering that our laboratory coronagraph was not optimized to reduce low order aberrations. We expect that a better optical design of the coronagraphic layout could improve substantially the contrast of the scheme.

%The setup is independent of the polarization of the input light.

%%%%%%%%%%%%%%%%%%%%%%%%%%%%%%%%%%%%%%%%%%%%%%%%%%%%%%%%%%%%%
\acknowledgments     %>>>> equivalent to \section*{ACKNOWLEDGMENTS}       

RE would like to thank DFG for support in the Priority Programme SPP 1385 in project NE 515 / 34-1 and the Abbe School of photonics for the Ph.D. grant.

%%%%%%%%%%%%%%%%%%%%%%%%%%%%%%%%%%%%%%%%%%%%%%%%%%%%%%%%%%%%%
%%%%% References %%%%%

\bibliography{report}   %>>>> bibliography data in report.bib

\begin{thebibliography}{10}

\bibitem{rod97}
{Roddier}, F. and {Roddier}, C., ``{Stellar Coronograph with Phase Mask},''
  {\em PASP}~{\bf 109},  815--820 (July 1997).

\bibitem{maw12}
{Mawet}, D., {Pueyo}, L., {Lawson}, P., {Mugnier}, L., {Traub}, W.,
  {Boccaletti}, A., {Trauger}, J.~T., {Gladysz}, S., {Serabyn}, E., {Milli},
  J., {Belikov}, R., {Kasper}, M., {Baudoz}, P., {Macintosh}, B., {Marois}, C.,
  {Oppenheimer}, B., {Barrett}, H., {Beuzit}, J.-L., {Devaney}, N., {Girard},
  J., {Guyon}, O., {Krist}, J., {Mennesson}, B., {Mouillet}, D., {Murakami},
  N., {Poyneer}, L., {Savransky}, D., {V{\'e}rinaud}, C., and {Wallace}, J.~K.,
  ``{Review of small-angle coronagraphic techniques in the wake of ground-based
  second-generation adaptive optics systems},'' in [{\em Society of
  Photo-Optical Instrumentation Engineers (SPIE) Conference
  Series}{\nolinebreak\hspace{0.1em}]},  {\em Society of Photo-Optical
  Instrumentation Engineers (SPIE) Conference Series} {\bf 8442} (Sept. 2012).

\bibitem{foo05}
{Foo}, G., {Palacios}, D.~M., and {Swartzlander}, Jr., G.~A., ``{Optical vortex
  coronagraph},'' {\em Optics Letters}~{\bf 30},  3308--3310 (Dec. 2005).

\bibitem{swa05}
{Swartzlander}, Jr., G.~A., ``{Broadband nulling of a vortex phase mask},''
  {\em Optics Letters}~{\bf 30},  2876--2878 (Nov. 2005).

\bibitem{maw05}
{Mawet}, D., {Riaud}, P., {Surdej}, J., and {Baudrand}, J., ``{Subwavelength
  surface-relief gratings for stellar coronagraphy},'' {\em Appl. Opt.}~{\bf
  44},  7313--7321 (Dec. 2005).

\bibitem{mal95}
{Malbet}, F., {Yu}, J.~W., and {Shao}, M., ``{High-Dynamic-Range Imaging Using
  a Deformable Mirror for Space Coronography},'' {\em Publications of the
  Astronomical Society of the Pacific}~{\bf 107},  386 (Apr. 1995).

\bibitem{maw10}
{Mawet}, D., {Pueyo}, L., {Moody}, D., {Krist}, J., and {Serabyn}, E., ``{The
  Vector Vortex Coronagraph: sensitivity to central obscuration, low-order
  aberrations, chromaticism, and polarization},'' in [{\em Society of
  Photo-Optical Instrumentation Engineers (SPIE) Conference
  Series}{\nolinebreak\hspace{0.1em}]},  {\em Society of Photo-Optical
  Instrumentation Engineers (SPIE) Conference Series} {\bf 7739} (July 2010).

\bibitem{Vasnetsov}
{Bazhenov}, V.~Y., {Soskin}, M.~S., and {Vasnetsov}, M.~V., ``{Screw
  Dislocations in Light Wavefronts},'' {\em Journal of Modern Optics}~{\bf 39},
   985--990 (May 1992).

\bibitem{swa09}
{Swartzlander}, Jr., G.~A., ``{The optical vortex coronagraph},'' {\em Journal
  of Optics A: Pure and Applied Optics}~{\bf 11},  094022 (Sept. 2009).

\bibitem{bez04}
{Bezuhanov}, K., {Dreischuh}, A., {Paulus}, G.~G., {Sch{\"a}tzel}, M.~G., and
  {Walther}, H., ``{Vortices in femtosecond laser fields},'' {\em Optics
  Letters}~{\bf 29},  1942--1944 (Aug. 2004).

\bibitem{err13}
{Errmann}, R., {Minardi}, S., and {Pertsch}, T., ``{A broad-band scalar vortex
  coronagraph},'' {\em MNRAS}~{\bf 435},  565--569 (Oct. 2013).

\bibitem{laf07}
{Lafreni{\`e}re}, D., {Marois}, C., {Doyon}, R., {Nadeau}, D., and {Artigau},
  {\'E}., ``{A New Algorithm for Point-Spread Function Subtraction in
  High-Contrast Imaging: A Demonstration with Angular Differential Imaging},''
  {\em The Astrophysical Journal}~{\bf 660},  770--780 (May 2007).

\end{thebibliography}
\bibliographystyle{spiebib}   %>>>> makes bibtex use spiebib.bst

\end{document}